\documentclass[9pt,twocolumn,twoside]{osajnl}

\journal{josab} 

\setboolean{shortarticle}{false} 

\ifthenelse{\boolean{shortarticle}}{\colorlet{color2}{color2b}}{\colorlet{color2}{color2}} 

\title{Characterization of frequency stability in EIT-based atomic clocks using\\ a differential detection scheme}

\author[1]{Melissa A. Guidry}
\author[2]{Elena Kuchina}
\author[1]{Irina Novikova}
\author[1,*]{Eugeniy E. Mikhailov}

\affil[1]{Department of Physics, College of William $\&$ Mary, Williamsburg, Virginia 23187, USA}
\affil[2]{Thomas Nelson Community College, Hampton, Virginia 23666, USA}

\affil[*]{Corresponding author: eemikh@wm.edu}

\dates{Compiled \today}

\ociscodes{
270.1670, Coherent optical effects;
020.1670, Coherent optical effects;
020.3690, Line shapes and shifts.
}
\doi{\url{http://dx.doi.org/10.1364/ao.XX.XXXXXX}}

\begin{abstract}
We investigate a recently proposed scheme for differential detection of the magneto-optical
rotation effect and its application to electromagnetically induced
  transparency (EIT) atomic clocks~\cite{SihongGu2015}. This scheme utilizes a
  linearly polarized bichromatic laser field that is EIT-resonant with
alkali atoms. The results of our study reveal that the suppression of the laser noise
can substantially improve the signal-to-noise ratio in EIT
atomic clocks. Our preliminary results demonstrate an order of magnitude improvement in clock stability
under some conditions when incorporating the differential detection scheme. 
\end{abstract}

\setboolean{displaycopyright}{true}

\begin{document}

\maketitle
\thispagestyle{fancy}

\ifthenelse{\boolean{shortarticle}}{\ifthenelse{\boolean{singlecolumn}}{\abscontentformatted}{\abscontent}}{}

\begin{figure}[h]
        \includegraphics[width=1.0\columnwidth]{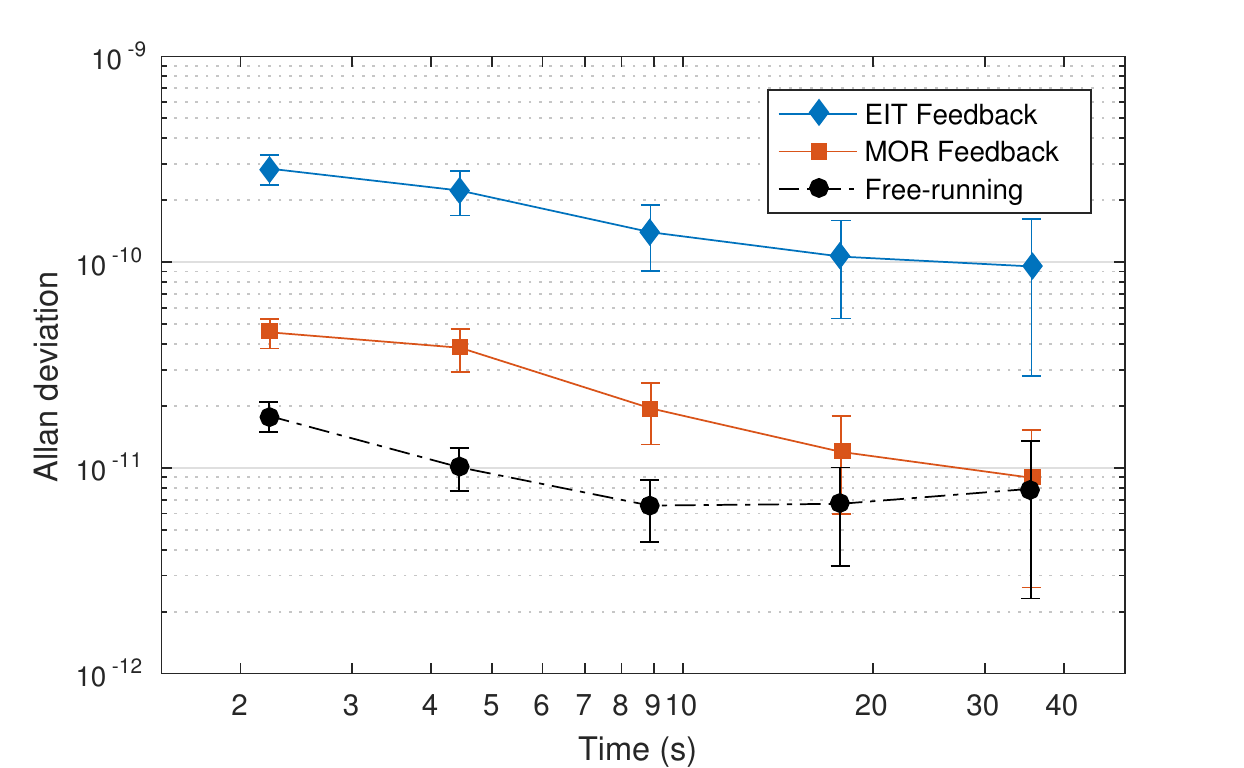}
        \caption{
                \label{fig:VCSEL}
                (Color online)
                Frequency stability measurements of a microwave crystal oscillator, locked to the EIT clock resonance in Rb vapor obtained with a current-modulated VCSEL and transmission EIT (blue diamonds) or differential MOR (red squares) detection. The VCSEL power was $75~\mu$W and the frequency counter gate time was $2$~s. 
		The intrinsic stability of the free-running crystal
		oscillator exceeded our measurement sensitivity; hence, its
		measured  stability is plotted for reference as the black
		dashed line with circles.
        }
\end{figure}

In the last few decades, all-optical measurements of electron spin in thermal ensembles of alkali metal atoms have been used as a basis for many precision measurement device prototypes,
such as atomic clocks~\cite{vanier_book} and magnetometers~\cite{romalis2007natphys,budker_optmagn_book}.
Many different interrogation schemes have been realized, and even more proposed. We focus on a relatively simple approach based on electromagnetically induced transparency (EIT)~\cite{FleischhauerRevModPhys05}, or in the language of~\cite{arimondo'96po,vanier05apb,Shah201021}, coherent population trapping (CPT); due to its simplicity, our measurement scheme is well suited for use in atomic clocks and magnetometers. 
 Traditionally, this method determines the frequency between hyperfine states of an alkali metal atom via the change in transmitted intensity of a bichromatic optical field.
 However, recently it has been proposed that a modified detection scheme sensitive to the changes in the light polarization due to magneto-optical rotation (MOR) may increase the performance of such frequency measurements~\cite{yanoIEEE2014,SihongGu2015,SunOE16}.
 In this paper, we experimentally compare two detection schemes (traditional intensity-based EIT and polarization-rotation-sensitive MOR).
 We demonstrate that in the case of a noisy laser, such as a vertical-cavity surface-emitting laser (VCSEL), the MOR scheme can significantly improve the stability of the frequency measurements, as shown in Fig.~\ref{fig:VCSEL}.

\begin{figure}[!htb]
	\includegraphics[width=\columnwidth]{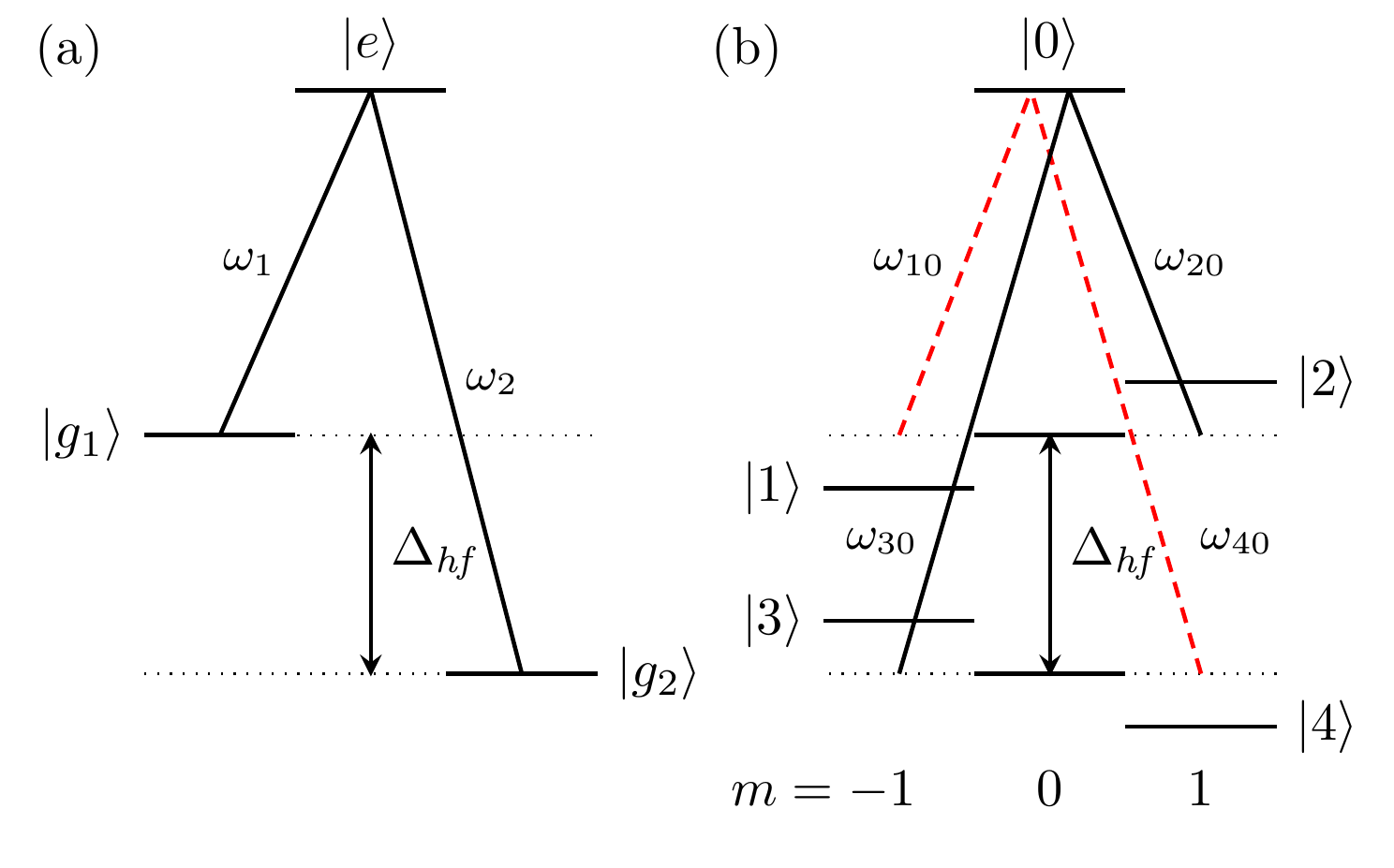}
\caption{(a) Simple three-level $\Lambda$ system, sufficient to explain the narrow electromagnetically induced transparency resonance. A dark state,  the quantum superposition of the ground states $|{g_1}\rangle$ and $|{g_2}\rangle$ decoupled from both optical fields, is produced under the conditions of the two-photon resonance when $\omega_1-\omega_2=\Delta_{\textit{hf}}$. (b) Double-$\Lambda$ configuration, formed for a linearly-polarized bichromatic laser field in the presence of a magnetic field. $|1\rangle$ and $|2\rangle$, and $|3\rangle$ and $|4\rangle$ are the $m=\pm 1$ Zeeman sublevels of the two hyperfine ground states, correspondingly, and $|0\rangle$ is the optical excited state. The two-photon resonance conditions in these two $\Lambda$ systems occur approximately at the same frequency as for the traditional magneto-insensitive $0$--$0$ ``clock'' transition, with the small difference arising from the gyromagnetic ratio difference between the two ground states.}
\label{fig:level_diagram}
\end{figure}

The improvement in performance results from the increased ability of differential MOR detection
to suppress common-mode intensity noise. As a brief reminder, in a traditional EIT system frequency measurements are determined by locating a peak transmission of a bichromatic optical field in a $\Lambda$ configuration, such that two components are in two-photon resonance with two hyperfine sublevels of the electron ground state (in our case, $5S_{1/2}\;F=1,2$ of ${}^{87}$Rb atoms, as shown in Fig.~\ref{fig:level_diagram}(a)).
 Under these conditions, the atoms are prepared in a non-interacting coherent superposition of the two ground states, known as a ``dark state''.
 If the frequency difference between the two optical fields (two-photon detuning) is swept across the frequency splitting between the two ground states, a narrow peak in optical transmission may be observed. The peak maximum matches the exact two-photon resonance conditions and the peak width is determined by the decoherence time of the dark state and the two-photon power broadening.
 Previous studies have reported EIT resonances as narrow as a few hundred Hz in cells with buffer gas~\cite{wynands'99,helm'01,XiaoMPL09}, and as narrow as a few Hz in cells with anti-relaxation wall coatings~\cite{budkerPRA05,romalisJAP09,balabasPRL10,BoudotAPL2015}.
 Such narrow transmission resonances make EIT attractive for miniature all-optical frequency standards (when the magnetic field-insensitive ``clock'' transition is probed)~\cite{vanier05apb,Shah201021,knappeOE05,knappeCM07} or magnetometers (when the Zeeman frequency shifts are measured).
 For such sensor applications, the two required optical fields are often generated by modulating the laser output at the hyperfine atomic frequency, and then stabilizing the microwave oscillator at the optical transmission peak.
 Under ideal conditions, the locked frequency precisely matches the frequency of the atomic transition $|{g_1}\rangle - |{g_2}\rangle$, and its stability improves with sharper contrast and narrower width of the EIT resonance.
 At the same time, technical noises (\textit{e.g.}, laser frequency noise and laser intensity noise) may introduce additional noises in the detection signal, thereby reducing sensor performance.
A number of methods have been introduced to reduce both systematic effects and technical noises caused by lasers, including time-resolved detection~\cite{zanonIEEE05,ClaironIEEE09,MitsunagaPRA13,Liu:13,kitchingPRA15}.

%

The proposed MOR effect occurs in the $lin||lin$ EIT configuration (which has been shown to produce high-contrast transmission resonances in ${}^{87}$Rb atoms) 
in the presence of an external longitudinal magnetic field,
 which is often applied to discriminate between transitions between various Zeeman sublevels~\cite{TaichenachevJETP2005,zibrovJETPLett05,BreschiPhysRevA.79.0638372009,ZibrovPhysRevA.81.013833.2010,mikhailov2010JOSAB_linparlin_clock}.
 This method takes advantage of steep dispersion, another popular EIT feature widely used in quantum information applications for group velocity manipulations~\cite{lukin03rmp,novikovaLPR12} and enhanced nonlinear magneto-optical effects~\cite{budkerRMP02}.
  The linearly polarized laser field may be decomposed into right and left circularly polarized optical fields. The clock transition is formed by a combination of two $\Lambda$ systems connecting the opposite $m=\pm1$ Zeeman sublevels, as shown in Fig.~\ref{fig:level_diagram}(b).
For a given $m$ Zeeman quantum number in two hyperfine states, the Zeeman shifts of the two levels are nearly equal in magnitude and opposite in sign.
 Under a stronger magnetic field,  the difference between the gyromagnetic ratios for the two hyperfine states results in a small difference in the exact two-photon resonant conditions of the two $\Lambda$ systems. Specifically, the hyperfine splitting angular frequency of $|1\rangle \rightarrow |4\rangle$ decreases
 while that of $|2\rangle \rightarrow |3\rangle$ increases by the same amount.
 Since the refractive index near the EIT peak is very sensitive to small two-photon detunings, this asymmetry results in a differential acquired phase between the two circular polarizations, equivalent to the rotation of the output linear polarization. 
Hence, when differential detection is performed using two outputs of an optical polarizer placed at $45^\circ$ with respect to the original polarization direction, a sharper resonant feature is expected when the modulation frequency is equal to the hyperfine splitting, $\omega_1 - \omega_2 = \Delta_{\textit{hf}}$.
 In addition, MOR detection schemes have demonstrated reduced intensity noise when compared to traditional EIT detection schemes  \cite{SihongGu2015,SunOE16}.
 Here, we show the first (to our best knowledge) experimental comparison of the frequency stabilities of EIT- and MOR-stabilized atomic clocks under different levels of laser noise and microwave oscillator noise.

\begin{figure}[h]
        \includegraphics[width=\columnwidth]{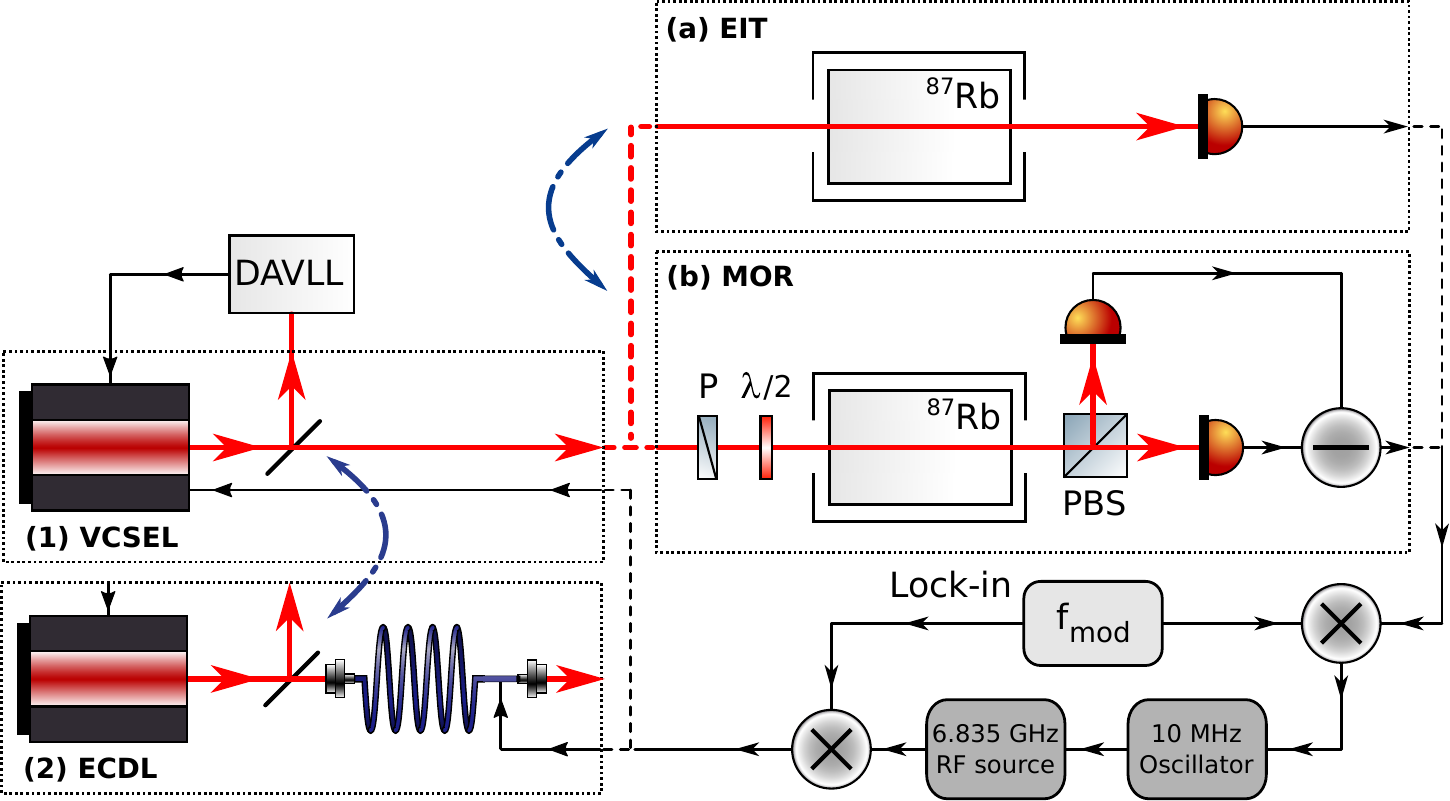}
        \caption{
                Schematic of the experimental setup. Left two blocks show interchangeable laser source: (1) a current-modulated VCSEL or (2) an ECDL with an external fiber modulator. The right blocks show two alternative detection methods: either (a) straight EIT transmission measurements or (b) differential MOR measurements.
 		Here P is a Glan laser polarizer, $\lambda/2$ is a half-wave plate, and PBS is a polarizing beam splitter.
        }
        \label{fig:MOR_setup}
\end{figure}

The experimental setup is shown in Fig.~\ref{fig:MOR_setup}, and more technical details are given in Ref.~\cite{mikhailov2009AJP_clock_for_undergrads}.
All the measurements are conducted at the D$_1$ line of ${}^{87}$Rb ($\lambda = 795$~nm).
To study the effect of laser noise on the clock stability of the two atomic detection schemes, we performed measurements  with two distinct diode lasers: a broad-band VCSEL laser (linewidth $\approx 300$~MHz) and a narrow-band external cavity diode laser (ECDL, linewidth $<100$~kHz).  
 Both lasers were modulated at the Rb hyperfine frequency to produce the two required frequency components:
the VCSEL was directly current-modulated, and the output of the ECDL was modulated using a fiber electro-optical modulator. 
In both cases, the laser carrier field was tuned to the $5 S_{1/2} F = 1 \rightarrow 5 P_{1/2} F' = 1$ transition, while the $+1$ modulation sideband frequency was tuned to the $5 S_{1/2} F = 2 \rightarrow 5 P_{1/2} F' = 1$ transition. 
The optical frequencies of either laser were stabilized using  a dichroic-atomic-vapor  laser lock  (DAVLL)~\cite{yashchukRSI00}. 
The modulation strength was adjusted to provide the $\approx 0.6$ ratio between the carrier and the first modulation sidebands, thereby reducing the effect of the light shifts~\cite{mikhailov2010JOSAB_linparlin_clock}.

The modulated linearly polarized laser beam (maximum power of $100~\mu$W) was directed into a cylindrical Pyrex cell (length $75$~mm; diameter $22$~mm)
 containing isotopically enriched ${}^{87}$Rb vapor and $30$~Torr of Ne buffer gas.
 The cell was mounted inside a three-layer magnetic shielding to reduce stray magnetic fields.
 The temperature was actively stabilized at $50^{\circ}$C, resulting in an atomic density of $1.5\times{10}^{11}$~cm$^{-3}$. 
 To lift the Zeeman degeneracy and isolate the magnetic field-insensitive resonance, we applied a weak homogeneous longitudinal magnetic field $B \approx 0.2$~G using a solenoid mounted inside the magnetic shielding.
The two schemes differed primarily in the detection method: when the clock was run using a traditional EIT configuration, a single photodiode (PD) was placed after the cell to measure the total transmitted power.
  When the clock was run using a differential MOR configuration, the output laser beam was incident on a polarizing beam splitter (PBS), rotated at $45^\circ$ with respect to the input polarization. The powers in the two PBS outputs were subtracted to obtain the differential signal. 

\begin{figure}[h]
        \includegraphics[width=0.5\textwidth]{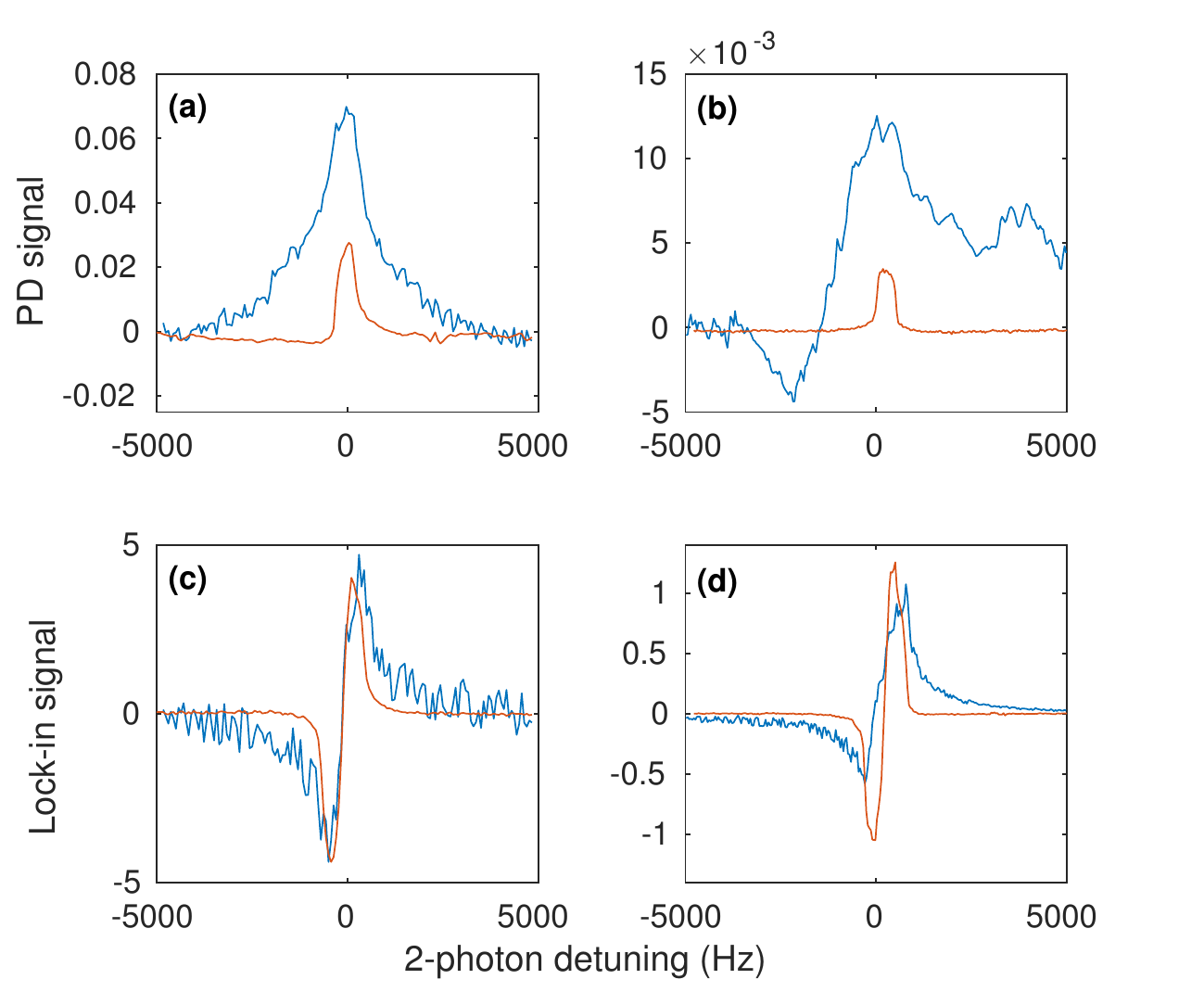}
        \caption{
		(Color online) EIT and MOR signal lineshapes (top row) and lock-in signals (bottom row) for the VCSEL (left) and the ECDL (right) using EIT (blue) or differential MOR (red) atomic feedback.
 The time constant of the lock-in is set to $0.1$ seconds.
 For the two measurements we used $35~\mu$W VCSEL power and $25~\mu$W of ECDL power, with base voltage levels on the PDs of $0.75$~V and $0.65$~V, respectively. The substantial low frequency noise of the EIT lineshape using the ECDL in plot (b) most likely arose from polarization instability in the optical fiber, resulting in slow intensity variation after the input polarizer.
        }
        \label{fig:lineshapes}
\end{figure}

To investigate the effect of the microwave oscillator phase noise on clock stability, we conducted the measurements using two different $10$~MHz oscillators. For a ``good'' oscillator, we used a high-quality low-noise 
 voltage-controlled oven-stabilized Wenzel 501-04609 crystal oscillator (CO), with a phase noise level of $-165$~dBc/Hz at a $10$~kHz offset. A Rigol DG1022 function generator served as a ``poor'' oscillator,  
 with nominal phase noise estimates of  $-108$~dBc/Hz at a $10$~kHz offset. The outputs of these oscillators were then multiplied up to the required $6.835$~GHz using a custom-designed phase locked loop circuit~\cite{mikhailov2010JOSAB_linparlin_clock}.
 Fine tuning is achieved via variable voltage of the $10$~MHz oscillator.


To lock the frequency of the microwave source to the transmission peak, a slow frequency modulation $f_{mod}=300$~Hz was superimposed on the $6.835$~GHz microwave modulation signal, and the final photodetector signal was then demodulated using a lock-in amplifier.
 The resulting error signal was used to correct the frequency of the $10$~MHz oscillator, thus locking the $6.835$~GHz source to the atomic resonance. 
 The frequency of the locked oscillator was measured by beating it with a reference $10$~MHz signal derived from a commercial atomic frequency standard (SRS FS725).

Fig.~\ref{fig:lineshapes} shows examples of EIT and MOR photodetector outputs and the corresponding demodulated error signals, using either the VCSEL or the ECDL, as the microwave modulation frequency (\textit{i.e.}, the two-photon detuning)
 was scanned around the microwave clock transition $6.835$~GHz. 
Using either laser, the linewidth of the MOR resonant feature is narrower than the corresponding EIT resonant feature (by a factor of $6$ in Fig.~\ref{fig:lineshapes}(a)).
 Moreover, the MOR signal demonstrates a significantly higher signal-to-noise ratio.
 Correspondingly, the lock-in signal for the MOR is steeper and less noisy given the same time constant. This improvement is evident when a broadband VCSEL is used, as in Fig.~\ref{fig:lineshapes}(c). In this case, the linewidth of the laser becomes comparable to the linewidth of the one-photon optical transition, hence we expect to see a significant FM-to-AM noise conversion in this case, compared to a much narrower ECDL~\cite{KitchingJOSA01}.
 The resulting intensity noise is shared between the left and right circular polarizations, such that it may be efficiently cancelled by the MOR differential detection scheme.  
Hence, we predict a significantly enhanced fractional stability of the oscillator when using the MOR signal, especially for VCSEL-based clocks.



The advantages of using a MOR detection scheme compared to a traditional EIT detection scheme for a noisy laser are clearly visible in Fig.~\ref{fig:VCSEL}, which shows the measured frequency stability for both methods using the VSCEL and the low noise crystal oscillator (CO).
As expected from the transmission signals in Fig.~\ref{fig:lineshapes}, the MOR signal improves clock stability over the EIT signal by a factor of $6$ for short averaging times ($\tau < 10\text{ s}$)  and by a factor of $10$ for longer times. 
The internal free-running stability of the high-quality crystal oscillator was initially a factor of two better than the MOR-feedback clock at short integration times, but was comparable at longer integration times, approaching the resolution of our frequency measurement system. 
Our short-term stability ($1$ s  $< \tau < 20$ s) measurements were most likely limited by the stability of our commercial reference clock SRS FS725 with manufacturer fractional stability $< 2 \times 10^{-11}$ at $1$ second.  
At longer integration times, the stability degraded due to uncontrolled temperature variations in the table-top apparatus.
Nevertheless, these measurements clearly demonstrate the superiority of MOR atomic clocks compared to traditional EIT atomic clocks, when a broadband laser is used.

\begin{figure}[t!]
        \includegraphics[width=1\columnwidth]{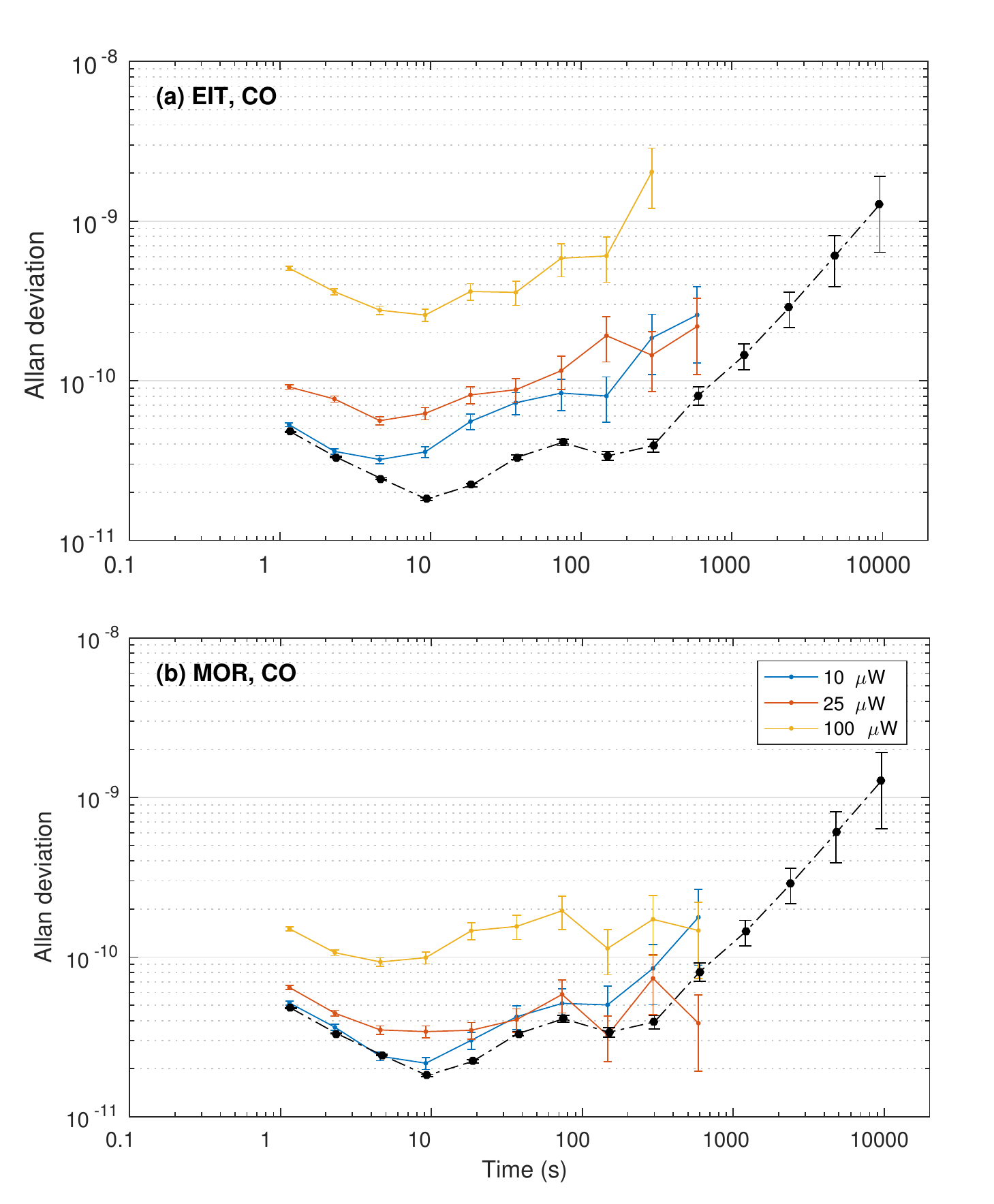}
        \caption{
                \label{fig:topt_CO}(Color online)
                Measurements of the fractional clock stability vs. averaging time at different powers of the ECDL and the low-noise crystal oscillator (CO) using either (a) EIT transmission or (b) differential MOR detection methods.
The frequency counter gate time is set to $1$ s.
The experimental stability limit (dotted black) is plotted for reference.
        }
\end{figure}

\begin{figure}[t!]
        \includegraphics[width=1\columnwidth]{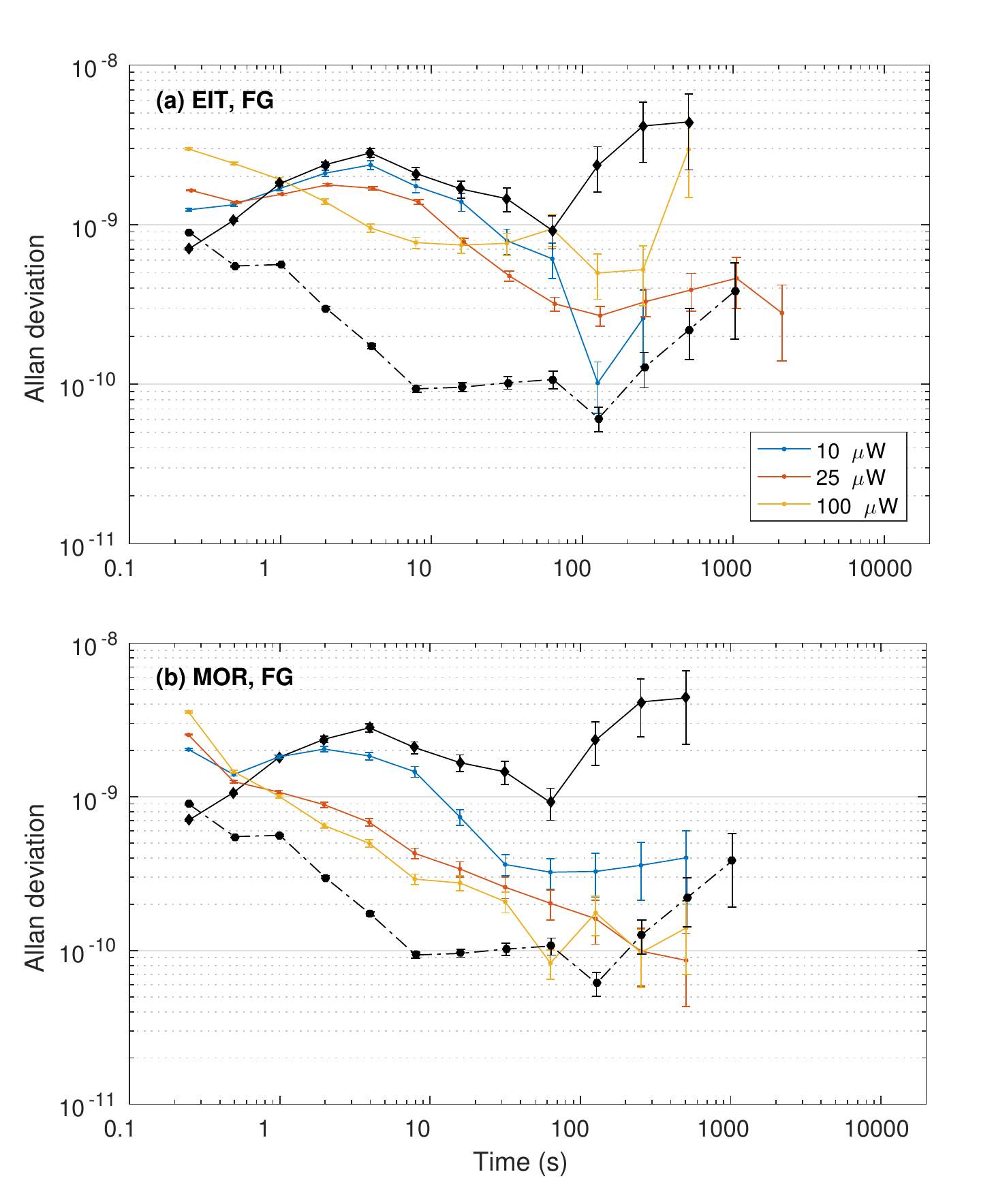}
        \caption{
                \label{fig:topt_FG}(Color online)
                Measurements of the fractional clock stability vs. averaging time at different powers of the ECDL and the noisy function generator (FG) using either (a) EIT transmission or (b) differential MOR detection methods.
The free-running function generator stability (solid black) and the measured crystal oscillator stability  (dashed black) are plotted for reference.
The frequency counter gate time is set to $0.1$ s.
        }
\end{figure}

A narrow-band ECDL was used to study the effects of the microwave oscillator noise on the clock stability for the two detection schemes.
 Fig.~\ref{fig:topt_CO} shows measured fractional Allan deviation for the high-quality low-noise crystal oscillator. As in the previous case, the stability of the free-running crystal oscillator exceeds the sensitivity of our frequency measurements, hence we use the free-running oscillator to estimate our stability measurement limitation for different gate settings.
 Even with the quieter ECDL, there is still some advantage in the use of MOR detection over EIT detection schemes: at higher powers, the short-term stability of the EIT clock [Fig.\ \ref{fig:topt_CO}(a)] is a factor of three worse than that of the MOR clock [Fig.\ \ref{fig:topt_CO}(b)]. For both EIT and MOR detection schemes, the stability generally improves at lower powers due to narrower resonances. However, under these conditions, the recorded sensitivity approached our measurement limits, making the performance comparison unreliable. 

These stability measurements were repeated using a much noisier microwave oscillator (a function generator, FG), as shown in Fig.~\ref{fig:topt_FG} for different laser powers. Locking the oscillator frequency to the atomic resonances  significantly improved its stability at time scales $> 1$~s with both EIT and MOR detection schemes. However, we observed only modest sensitivity improvement switching from EIT to MOR measurements. 
Unfortunately, we were unable to measure the stability of the atomic clocks with both a noisy oscillator and a noisy laser, as the VCSEL stopped functioning.
 
 \begin{figure}[t]
	\includegraphics[width=1.0\columnwidth]{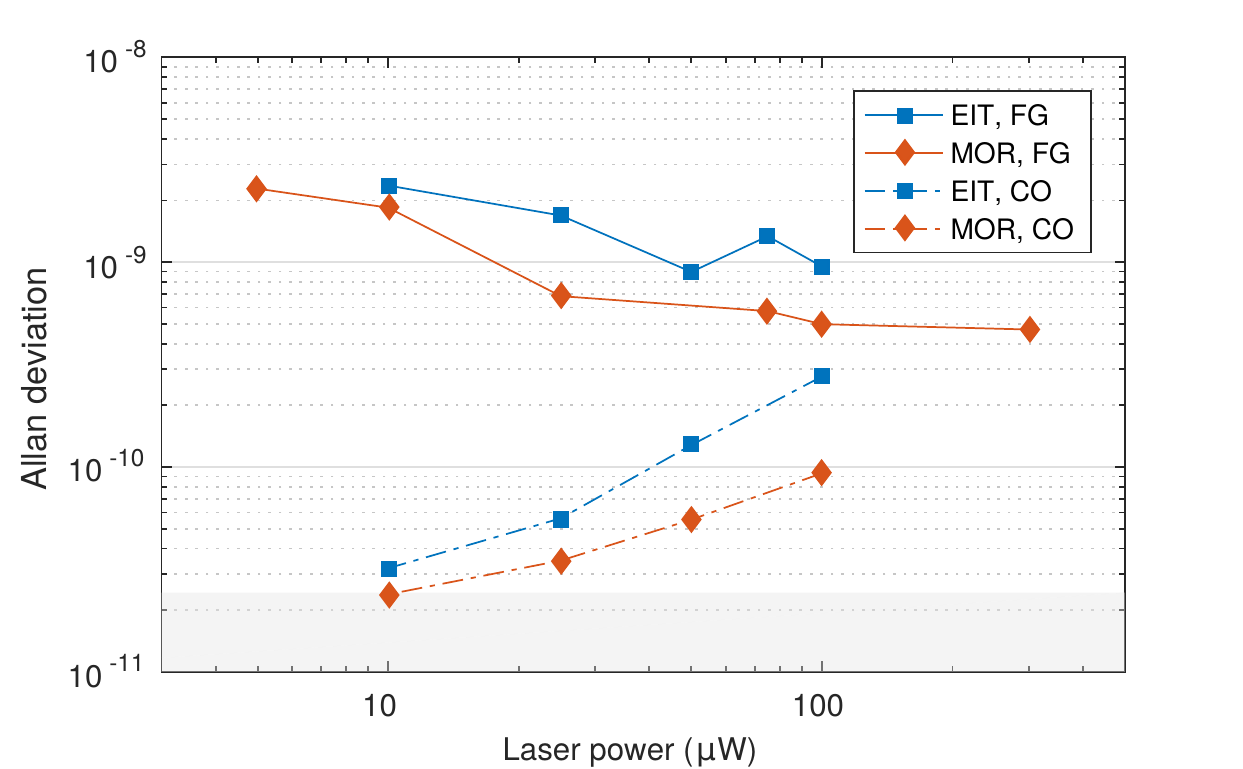}
	\caption{
		\label{fig:allan_vs_power}(Color online)
		Dependence of the Allan deviation measured at a $4.3$ s averaging time as a function of laser power.
		The shaded region represents the stability limit of the experiment, measured as the performance of the free-running crystal oscillator at that integration time.
		The error bars on each measurement are smaller than the marker size.
	}
\end{figure}

To better illustrate these observations, Fig.~\ref{fig:allan_vs_power} shows the measured fractional Allan deviation of the clock frequency using different setups at an averaging time of $4.3$~seconds with different power levels.
 The stability of the MOR-locked clock was always greater than that of the EIT-locked clock, but the improvement was laser power dependent: for a weak laser beam  we achieved the smallest fractional Allan deviation, but the observed gain of MOR over EIT was modest, potentially due to the measurement limitations.
 At higher laser power, the overall frequency deviation increased by nearly a factor of ten, but the MOR-locked clock stability was a factor of two greater than the EIT-locked clock stability.
This degradation with increased power is most likely due to power broadening of the EIT and MOR signals. 
Using the function generator, the gains of the MOR setup over the EIT setup are similarly modest over all powers; however, atomic clock stability improves with increased laser power.
The function generator has large phase noise, such that the EIT and MOR signals are quite broad, reducing the effect of the power broadening.
Hence, an increase in power will improve the signal-to-noise ratio without a significant increase in signal width due to power broadening.

We can qualitatively explain the lack of significant improvement if the main source of fluctuations comes from the microwave source, rather than the laser.
 As discussed above, the phase noise of the laser is transformed into the amplitude noise mainly via the single-photon optical transmission lineshape.
 Thus, added intensity fluctuations are mostly identical for both circular polarizations, and are efficiently subtracted in MOR differential detection, even though they directly affect the EIT clock performance.
 However, the microwave source adds noise into the two-photon detuning, resulting in broadening of both EIT and MOR signals in a similar manner and reducing the difference in performance between the two schemes. 
The MOR signal resonances tend to be narrower then EIT signal resonances (see Fig.~\ref{fig:lineshapes}), hence the MOR detection scheme facilitated better clock stability under otherwise identical conditions. 

In summary, we systematically studied the frequency stability of EIT-based atomic clocks that used either pure transmission or a differential MOR signal for feedback.
 We investigated the effect of the laser linewidth (\textit{i.e}. fluctuations in the one-photon detuning), microwave source noise (\textit{i.e.}, fluctuations in the two-photon detuning), and laser power on clock stability.
 Under all experimental conditions, the MOR scheme exhibited enhanced fractional stability over the EIT scheme. 
For a narrow-band ECDL the gains were modest (around a factor of two at best), even while using a very noisy microwave oscillator.
The greatest improvement using the MOR signal was seen when the laser itself was very noisy, as the differential detection scheme allowed cancellation of the intensity noise caused by laser frequency fluctuations. 
A VCSEL-driven miniaturized atomic clock may enjoy significant improvement in short-term frequency stability when employed with a differential MOR detection scheme, with little additional increase in power consumption, volume, or experimental complexity. 

The authors would like to thank M. Kopreski for valuable discussion. M. A. Guidry was supported by the Virginia Space Grant Consortium Undergraduate Research Scholarship.

\end{document}